\documentclass[conference]{IEEEtran}
\IEEEoverridecommandlockouts

\usepackage{graphicx}

\usepackage{amsmath}
\usepackage{amsfonts}

\usepackage{cite}
\usepackage{multirow}
\usepackage{multicol}
\usepackage[linesnumbered,ruled,vlined, lined, commentsnumbered, longend]{algorithm2e} 
\usepackage{amsmath,amssymb,amsfonts}
\usepackage{algorithmic}
\usepackage{graphicx}
\usepackage{textcomp}
\usepackage{xcolor}

\usepackage{cite}
\usepackage{multirow}
\usepackage{multicol}

\usepackage[linesnumbered,ruled,vlined, lined, commentsnumbered, longend]{algorithm2e} 

\usepackage{etoolbox}

\usepackage{amsmath,amssymb,amsfonts}

\usepackage{xcolor}
\usepackage[colorlinks = true,
            linkcolor = blue,
            urlcolor  = blue,
            citecolor = blue,
            anchorcolor = blue]{hyperref}
\usepackage{orcidlink}

\usepackage{scalerel}
\usepackage{tikz}
\usetikzlibrary{svg.path}
\usepackage{scalerel}[2016/12/29]

\definecolor{orcidlogocol}{HTML}{A6CE39}

\tikzset{
  orcidlogo/.pic={
    \fill[orcidlogocol] svg{M256,128c0,70.7-57.3,128-128,128C57.3,256,0,198.7,0,128C0,57.3,57.3,0,128,0C198.7,0,256,57.3,256,128z};
    \fill[white] svg{M86.3,186.2H70.9V79.1h15.4v48.4V186.2z}
                 svg{M108.9,79.1h41.6c39.6,0,57,28.3,57,53.6c0,27.5-21.5,53.6-56.8,53.6h-41.8V79.1z M124.3,172.4h24.5c34.9,0,42.9-26.5,42.9-39.7c0-21.5-13.7-39.7-43.7-39.7h-23.7V172.4z}
                 svg{M88.7,56.8c0,5.5-4.5,10.1-10.1,10.1c-5.6,0-10.1-4.6-10.1-10.1c0-5.6,4.5-10.1,10.1-10.1C84.2,46.7,88.7,51.3,88.7,56.8z};
  }
}

\newcommand\orcid[1]{\href{https://orcid.org/#1}{\mbox{\scalerel*{
\begin{tikzpicture}[yscale=-1,transform shape]
\pic{orcidlogo};
\end{tikzpicture}
}{|}}}}

\def\BibTeX{{\rm B\kern-.05em{\sc i\kern-.025em b}\kern-.08em
    T\kern-.1667em\lower.7ex\hbox{E}\kern-.125emX}}
\begin{document}

\title{Collective Learning Mechanism based Optimal Transport Generative Adversarial Network for Non-parallel Voice Conversion
}

\author{\IEEEauthorblockN{Sandipan Dhar$^1$, Md. Tousin Akhter$^2$, Nanda Dulal Jana$^1$, Swagatam Das$^3$}
    \IEEEauthorblockA{$^1$\textit{Department of Computer Science and Engineering, National Institute of Technology Durgapur, India.}}
    \IEEEauthorblockA{$^2$\textit{Department of Computer Science and Engineering, Indian Institute of Technology Bombay, India.}}
    \IEEEauthorblockA{$^3$\textit{Electronics and Communication Sciences Unit, Indian Statistical Institute, Kolkata, India.}}
    \IEEEauthorblockA{Email: \url{sandipandhartsk03@gmail.com}, \url{tousin@cse.iitb.ac.in}, \url{ndjana.cse@nitdgp.ac.in}, \url{swagatam.das@isical.ac.in}}}

\maketitle

\begin{abstract}
After demonstrating significant success in image synthesis, Generative Adversarial Network (GAN) models have likewise made significant progress in the field of speech synthesis, leveraging their capacity to adapt the precise distribution of target data through adversarial learning processes. Notably, in the realm of State-Of-The-Art (SOTA) GAN-based Voice Conversion (VC) models, there exists a substantial disparity in naturalness between real and GAN-generated speech samples. Furthermore, while many GAN models currently operate on a single generator discriminator learning approach, optimizing target data distribution is more effectively achievable through a single generator multi-discriminator learning scheme. Hence, this study introduces a novel GAN model named Collective Learning Mechanism-based Optimal Transport GAN (CLOT-GAN) model, incorporating multiple discriminators, including the Deep Convolutional Neural Network (DCNN) model, Vision Transformer (ViT), and conformer. The objective of integrating various discriminators lies in their ability to comprehend the formant distribution of mel-spectrograms, facilitated by a collective learning mechanism. Simultaneously, the inclusion of Optimal Transport (OT) loss aims to precisely bridge the gap between the source and target data distribution, employing the principles of OT theory. The experimental validation on VCC 2018, VCTK, and CMU-Arctic datasets confirms that the CLOT-GAN-VC model outperforms existing VC models in objective and subjective assessments.
\end{abstract}


\section{Introduction}
Voice Conversion (VC) is the process of altering a source speaker's vocal identity to match that of a target speaker while retaining the source speaker's original linguistic content \cite{Survey, VC-Survey}. In the early developments of VC research, Gaussian Mixture Models (GMMs) \cite{GMM}, Phonetic Posteriorgrams (PPGs) \cite{PPG}, etc., \cite{More, More-1, HMM, VQ-VAE} were widely used. However, over the last few years, due to the overwhelming data generation ability of the Generative Adversarial Network (GAN) model \cite{gan2014}, it has become an effective alternative for performing the VC task. In one-to-one VC, models such as ALGAN-VC\cite{ALGAN} have demonstrated notable enhancements in performance through implementing adaptive learning mechanisms. Recent variants of GAN-based VC models \cite{gan5,IJCNN-1,IJCNN-2}, like Mask-CycleGAN-VC \cite{Maskcyclegan-VC}, FLSGAN-VC \cite{Flsgan} and FID-RPRGAN-VC \cite{FID-RPR-GAN-VC}, have improved the performance of non-parallel VC significantly. Both FLSGAN-VC \cite{Flsgan} and  FID-RPRGAN-VC \cite{FID-RPR-GAN-VC} generates target mel-spectrograms by considering feature specific evaluation metrics as loss functions to address the over-smoothing problem. Similar advancements have been made in the field of many-to-many VC \cite{GLGAN-VC}. However, State-Of-The-Art (SOTA) one-to-one GAN-based VC models, such as FID-RPRGAN-VC \cite{FID-RPR-GAN-VC}, FLSGAN-VC \cite{Flsgan}, etc., still show a notable gap in naturalness between real and synthesized speech. 
\par
Most of these GAN-based VC models are developed using Convolutional Neural Network (CNN) model-based discriminator architectures that capture the feature information from mel-spectrograms by learning spatial hierarchies of features, ranging from low to high-level patterns \cite{CNN_SR}. However, in recent years, Vision Transformer (ViT) \cite{ViT-GAN} has shown significant performance for Speaker Identification (SI) tasks \cite{ViT-CNN} because it obtains the information of local feature distribution more precisely from small patches of input data by utilizing the concept of attention mechanism. In this context, the conformer \cite{Conformer-GAN} models have also shown substantial performance improvement for SI and Speech Command Recognition (SCR) tasks. In GAN models, the discriminator is similar to speaker verification models, as it determines whether a given sample belongs to the real class or not. It can also be viewed as a SI task. 
\par
Therefore, it is worthwhile to integrate ViT and conformer models with CNNs by leveraging the potential benefits of all in a single GAN-based VC framework following a collaborative learning approach, particularly within a single-generator, multi-discriminator setup. This approach would enable more detailed feedback to the generator, with each discriminator concentrating on different aspects of formant distribution or mel-spectrogram features, potentially resulting in more robust and precise VC. From the existing literature \cite{Survey}, it is observed that the preceding GAN-based VC models were developed based on different probability divergence measures as loss functions (e.g., $L_2$ loss, cross-entropy loss) \cite{KL-JS-1, ALGAN, KL-JS}. Yet, there is an opportunity to investigate Optimal Transport (OT) metrics, as described in \cite{Sinkhorn-Divergence} as loss functions. These metrics consider the minimal work needed to transform one probability distribution into another instead of solely focusing on their values at individual points. These metric losses have several advantages over KL and JS divergence \cite{Wasserstein-Loss}. 
\par
This study introduces a collective learning mechanism-based optimal transport GAN model for non-parallel VC, termed as CLOT-GAN-VC, which employs a single generator and multiple discriminators by combining the effectiveness of Deep CNN (DCNN), ViT, and conformer architectures within a single framework. This novel GAN model integrates a collaborative learning mechanism and an OT loss function. The underlying Sinkhorn algorithm in OT loss helps to measure the difference between source and target feature distribution in an efficient manner. Unlike the traditional ensemble learning approach, the proposed collective learning approach allows each discriminator to update its weights based on its current evaluation of feature similarity between the generated mel-spectrogram and its original counterpart. This weighted averaging of individual decisions provides the generator with a comprehensive, multidimensional perspective of its generated samples (mel-spectrograms), with each discriminator simultaneously learning from its respective loss. As a result, the entire system learns through a collective learning approach involving $n$ discriminators. This work is highly motivated by the concept of multi-agent systems \cite{Multi-agent} and multiplayer games \cite{Multi-player-game}. The key contributions of this work are summarized as follows:
\begin{enumerate}
\item{We present an innovative GAN model employing a single generator and multiple discriminators. These discriminators utilize DCNN, ViT, and Conformer architecture. Their collaboration is designed to capture the formant distribution within mel-spectrograms effectively.}
\item{We introduce a collective learning mechanism to this work to update the multiple discriminators' learnable parameters. }
\item{We utilize the concept of OT loss to trace
the difference between source and target feature distribution more efficiently based on the concept of OT theory.}
\end{enumerate}

\par
We validate our model's effectiveness using the VCC 2018, CSTR-VCTK, and CMU Arctic speech datasets, assessing performance with Mel-Cepstral Distortion (MCD), Modulation Spectra Distance (MSD), and Mean Opinion Score (MOS). Our results demonstrate that the speech samples generated by our proposed model show improved quality and speaker similarity compared to the SOTA VC models like MaskCycleGAN-VC \cite{Maskcyclegan-VC} and MelGAN-VC \cite{MelGAN-VC}.
\par
The subsequent sections of the paper are structured in the following manner. Section \ref{related-work} briefly discusses the related studies and their findings. Section \ref{proposed-approach} provides a comprehensive explanation of the CLOT-GAN-VC model. Section \ref{exp-setup} includes the information about the training details and the speech datasets employed for performance evaluation. In Section \ref{results}, results are provided, and the outcome of the experiments is thoroughly discussed. Finally, Section \ref{conclusion} concludes the paper and highlights some interesting prospects for future research in GAN-based VC.
\section{Related Work} \label{related-work}
Among the recent works of VC, a significant number of publications are based on different variants of GAN models \cite{Survey}. Therefore, the literature survey of this work is primarily focused on one-to-one GAN-based non-parallel VC. In Cyclegean \cite{CycleGEAN}, the authors proposed a multiple discriminator-based GAN model to trace the speaker similarity between the original and the generated speech samples, rather than using a single discriminator to classify the generated samples as real or fake. This is an extension of the CycleGAN-VC \cite{gan5} model. In \cite{Maskcyclegan-VC}, Kaneko et al. proposed MaskCycleGAN-VC model where the CycleGAN-VC2 framework \cite{CycleGAN-VC2} is considered for mel-spectrogram-based one-to-one VC. Their proposed model is developed based on the concept of filling the frames \cite{Filling-black} and obtained better results than the preceding models of the CycleGAN-VC family. In \cite{Flsgan}, the authors proposed  FLSGAN-VC model for the mel-spectrogram-based style transfer of vocal features, from source to target speaker. Here, the source mel-spectrogram is divided into multiple frames and the target mel-spectrogram is generated using multiple generators (the same number of generators as the number of frames). Moreover, the siamese network is also incorporated in \cite{Flsgan} for calculating the siamese loss. In a similar way, FID-RPRGAN-VC \cite{FID-RPR-GAN-VC} is trained with a Frechet Inception Distance (FID)-based loss function, utilizing two DCNN discriminators. However, the Inception V3 model is used exclusively for computing the FID score. Hence, the preceding literature indicates that there is potential for more efficient target distribution acquisition by exploring single-generator multi-discriminator learning strategies considering different deep learning architectures.

\section{Proposed approach} \label{proposed-approach}
\begin{figure}[t]
    \centering
    \includegraphics[height=7cm, width=8.45cm]{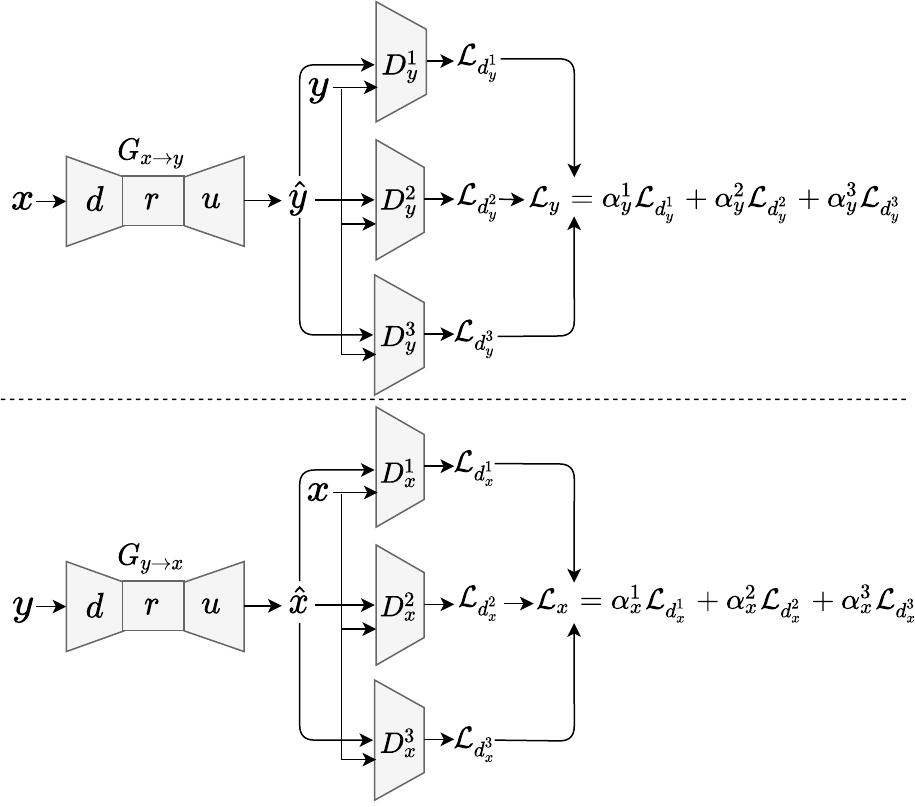}
    \caption{Overview of the proposed CLOT-GAN-VC model}
    \label{fig:tf-gan}
\end{figure}

This section briefly discusses the proposed CLOT-GAN-VC model and its training mechanism. As shown in Fig. \ref{fig:tf-gan}, 
$\mathrm{x\in X}$ and $\mathrm{y\in Y}$ are the mel-spectrograms of the source and target speakers, respectively, for the generator $G_{x\rightarrow y}$, and vice-versa. In Fig. \ref{fig:tf-gan}, $d$ implies the downsample block, $r$ implies the residual block, and $u$ implies the upsample block of the generator. The architectural framework of the CLOT-GAN-VC model is developed keeping the framework of Mask-CycleGAN-VC \cite{Maskcyclegan-VC} as the backbone. The generated mel-spectrograms $\hat{x}$ and $\hat{y}$ are fed to each of the discriminators (i.e. $D^{k=1\hspace{0.1cm}to\hspace{0.1cm}n}_{x}$ and $D^{k=1\hspace{0.1cm}to\hspace{0.1cm}n}_{y}$, here $k$ indicates the identity of the discriminators that varies from $1$ to $n$) as shown in Fig. \ref{fig:tf-gan}. This work considers three different types of discriminator architectures (i.e., $n$ is considered $3$). The three discriminators considered in this work are DCNN \cite{Maskcyclegan-VC}, ViT \cite{ViT-GAN}, and conformer \cite{Conformer-GAN} architectures. The architectural framework of the generator and the DCNN discriminator are kept the same as \cite{Maskcyclegan-VC}. The ViT and conformer architecture are adapted based on \cite{ViT-GAN} and \cite{Conformer-GAN}, respectively. The multiple discriminators incorporated in this work capture different aspects of feature information from the mel-spectrograms (such as deep convolutional features utilizing DCNN, patch-wise transformer encoded features utilizing ViT, and conformer extracted local-global features). Thus, the different feature embeddings extracted by each of the considered frameworks capture various nuances of the input mel-spectrogram, including both global and local distribution-related information. 
\par
Therefore, we introduce a collective learning mechanism to collaboratively learn the feature distribution of the mel-spectrograms (while mapping the mel-spectrograms from source to target) by utilizing multiple discriminators (rather than relying on a single model). The collective learning mechanism is presented in Algorithm \ref{algo_3}.
\LinesNumberedHidden
\begin{algorithm}[t]
\caption{Collective Learning Mechanism}
\label{algo_3}
\LinesNumberedHidden
\KwIn{$x \in X$, $y \in Y$}
\KwOut{Optimal parameters ${\boldsymbol{\theta}}_d^*$}
\KwData{Discriminator OT loss: $\mathcal{L}_{{d}^{k}_{c}}$ ($c \in \{x, y\}$, $k \in \{1, 2, ..., n\}$), Optimizer: Opt, Iterators: $i=0$ and $j=0$, Learnable parameters: ${\boldsymbol{\theta}_{d}} \in \{{\theta}^{k=1}_{{d_x}}, {\theta}^{k=2}_{{d_x}}, ..., {\theta}^{k=n}_{{d_x}}, {\theta}^{k=1}_{{d_y}}, {\theta}^{k=2}_{{d_y}}, ..., {\theta}^{k=n}_{{d_y}}\}$}
\While{$\boldsymbol{\theta}_{d}$ not converged}{
    $\mathcal{L}_{{d}^{k=1}_{y_{i}}}, \mathcal{L}_{{d}^{k=2}_{y_{i}}}, ..., \mathcal{L}_{{d}^{k=n}_{y_{i}}} \gets \mathcal{\textit{CLOT-GAN-VC}}(x, y, {\theta}^{k=1}_{{d_y}_{i}}, {\theta}^{k=2}_{{d_y}_{i}}, ..., {\theta}^{k=n}_{{d_y}_{i}})$\;
    $\mathcal{L}_{{tot}_{y_{i}}} = \sum_{k=1}^{n} \mathcal{L}_{{d}^{k}_{y_{i}}}$\;
    $\alpha^{k}_{y_{i}} = \frac{\mathcal{L}_{{tot}_{y_i}} - \mathcal{L}_{{d}^{k}_{y_i}}}{\mathcal{L}_{{tot}_{y_i}}}$\;
    $\mathcal{L}_{{y_i}} = \sum_{k=1}^{n} \alpha^{k}_{y_i} \mathcal{L}_{{d}^{k}_{y_i}}$\;
    $i = i + 1$\;
    Update ${\theta}^{k=1}_{{d_y}_i}, {\theta}^{k=2}_{{d_y}_i}, ..., {\theta}^{k=n}_{{d_y}_i}$ based on $\mathcal{L}_{{y_i}}$ using Opt\;
    
    $\mathcal{L}_{{d}^{k=1}_{x_j}}, \mathcal{L}_{{d}^{k=2}_{x_j}}, ..., \mathcal{L}_{{d}^{k=n}_{x_j}} \gets \mathcal{\textit{CLOT-GAN-VC}}(y, x, {\theta}^{k=1}_{{d_x}_j}, {\theta}^{k=2}_{{d_x}_j}, ..., {\theta}^{k=n}_{{d_x}_j})$\;
    $\mathcal{L}_{{tot}_{x_j}} = \sum_{k=1}^{n} \mathcal{L}_{{d}^{k}_{x_j}}$\;
    $\alpha^{k}_{x_j} = \frac{\mathcal{L}_{{tot}_{x_j}} - \mathcal{L}_{{d}^{k}_{x_j}}}{\mathcal{L}_{{tot}_{x_j}}}$\;
    $\mathcal{L}_{{x_j}} = \sum_{k=1}^{n} \alpha^{k}_{x_j} \mathcal{L}_{{d}^{k}_{x_j}}$\;
    $j = j + 1$\;
    Update ${\theta}^{k=1}_{{d_x}_j}, {\theta}^{k=2}_{{d_x}_j}, ..., {\theta}^{k=n}_{{d_x}_j}$ based on $\mathcal{L}_{{x_j}}$ using Opt\;
}
\textbf{return} optimal parameters ${\boldsymbol{\theta}}_d^*$\;
\end{algorithm}
As per Algorithm \ref{algo_3}, the input mel-spectrograms $x$ and $y$ (i.e. source $x$ and target $y$ for $G_{x\rightarrow y}$, and vice-versa) and the corresponding learnable parameters $\boldsymbol{\theta}_{d}$ belonging to $n$ number of discriminators (i.e. ${\theta}^{k=1\hspace{0.1cm}to\hspace{0.1cm}n}_{{d_y}}\in D^{k=1\hspace{0.1cm}to\hspace{0.1cm}n}_{y}$, and vice-versa) are provided as the input to the CLOT-GAN-VC model. Thereafter, the discriminator loss for each of the corresponding discriminators ($\mathcal{L}_{{d}^{k=1\hspace{0.1cm}to\hspace{0.1cm}n}_{c{}}} \in D^{k=1\hspace{0.1cm}to\hspace{0.1cm}n}_{c}$, here $c$ implies the speaker class $x$ and $y$) are obtained in each training epochs, as presented in Algorithm \ref{algo_3} and Fig. \ref{fig:tf-gan}. The discriminator loss considered in this work is the OT loss function. After that, the total loss $\mathcal{L}_{{tot}_x}$ and $\mathcal{L}_{{tot}_y}$ are obtained by summing up all the corresponding discriminator losses. To find the participation or contribution of each of the discriminators (in terms of loss) in the total loss, the participation weights ${\alpha}^{k=1\hspace{0.1cm}to\hspace{0.1cm}n}_{x}$ and ${\alpha}^{k=1\hspace{0.1cm}to\hspace{0.1cm}n}_{y}$  are calculated for the respective discriminators. As per the Algorithm \ref{algo_3}, the highest participation weight is associated with the least discriminator OT loss (i.e., $\alpha \propto \frac{1}{\mathcal{L}_d}$). Afterward, as provided in Algorithm \ref{algo_3}, each of the participation weights is multiplied to the corresponding discriminator losses for obtaining the final total discriminator loss $\mathcal{L}_{x}$ and  $\mathcal{L}_{y}$ respectively. As illustrated in Algorithm \ref{algo_3}, throughout the training process, the optimizer updates the values of the learnable parameters belonging to the respective discriminators based on $\mathcal{L}_{x}$ and  $\mathcal{L}_{y}$ for each training epochs. 
\par
\textbf{Discriminator OT loss:}
In the discriminator OT loss of our proposed model, the concept of OT theory \cite{Sinkhorn} is used to gauge the difference between two probability distributions by minimizing their optimal transport distance $\mathcal{W}_{c}$. This distance is the trace of the product between the optimal transport cost matrix $C\in\mathbb{R}^{N\times N}$ and soft matchings matrix $M\in\mathbb{R}^{N\times N}$ (here, $M$ is obtained using the sinkhorn algorithm, and $N$ is considered as $4$) as expressed in Eq. (\ref{eq:OT-1}) and  Eq. (\ref{eq:OT-2}),
\begin{equation}
\begin{split}
\label{eq:OT-1}
\mathcal{W}_{c}(X,Y) = \inf_{M \in \mathcal{M}} \operatorname{Tr}[M C^{T}],
\end{split}
\end{equation}
\vspace{-0.5cm}
\begin{equation}
\begin{split}
\label{eq:OT-2}
{C}_{p,q}={c}({\textbf{x}_p,\textbf{y}_q}).
\end{split}
\end{equation}
As depicted in Equation (\ref{eq:OT-2}), $C$ is the cost associated with transporting the $p^{th}$ data vector $\textbf{x}_{p}$ in mini-batch $X$ to the $q^{th}$ data vector $\textbf{y}_q$ in mini-batch $Y$. This work defines the cost function $c$ as the cosine distance \cite{Sinkhorn}. The mathematical representation of the discriminator OT loss $\mathcal{L}_{d}$ is provided in Equation (\ref{eq:OT-3}),
\begin{equation}
\begin{split}
\label{eq:OT-3}
\mathcal{L}_{d}=\mathcal{W}_{c}(X,X^{'})+\mathcal{W}_{c}(X,Y^{'})+\mathcal{W}_{c}(X^{'},Y)+\\
\mathcal{W}_{c}(X^{'},Y^{'})-2\mathcal{W}_{c}(X,X^{'})-2\mathcal{W}_{c}(Y,Y^{'}),
\end{split}
\end{equation}
where $X$ and $X^{'}$ represent mini-batches independently sampled from the distribution of real class, while $Y$ and $Y^{'}$ denote independent mini-batches sampled from the distribution of generated class. To calculate the discriminator OT loss $\mathcal{L}_{d}$, in this work, the last layer flatten vector \cite{Flatten} of the respective discriminators are considered. These flattened vectors act as feature representations, preserving key speech characteristics in the latent space as feature embeddings, which vary based on the model architecture. 
\par
By framing the vocal style transfer task (associated with VC) as an optimal transport problem, the primary goal of applying OT theory in our work is to make it easier to track the distribution shift toward the desired optimal state. A generated distribution that accurately replicates the target distribution can thus be considered a better solution for the VC task. 
\par
\textbf{Generator loss:} The generators learn the target distributions by minimizing the least squares loss $\mathcal{L}^{x\rightarrow y}_{g}$ and $\mathcal{L}^{y\rightarrow x}_{g}$ as given in Eq. (\ref{eq:Equation07}) and Eq. (\ref{eq:Equation08}),
\begin{equation}
    \mathcal{L}^{x\rightarrow y}_{g}= \Sigma^{k=n}_{k=1}\frac{1}{2}\mathbb{E}_{x\sim P_x}\beta^{k}_{y}(D^{k}_{y}(G_{x\rightarrow y}(x))-1)^2,
    \label{eq:Equation07}
\end{equation}
\vspace{-0.4cm}
\begin{equation}
    \mathcal{L}^{y\rightarrow x}_{g}= \Sigma^{k=n}_{k=1}\frac{1}{2}\mathbb{E}_{y\sim P_y}\beta^{k}_{x}(D^{k}_{x}(G_{y\rightarrow x}(y))-1)^2.
    \label{eq:Equation08}
\end{equation}
As shown in Eq. (\ref{eq:Equation07}) and Eq. (\ref{eq:Equation08}), the generator loss is also obtained based on the collective learning mechanism of Algorithm \ref{algo_3} and the learnable parameters of the generator (i.e., by replacing $\boldsymbol{\theta}_{d}$ in Algorithm \ref{algo_3} with generator's learnable parameters $\boldsymbol{\theta}_{g}$) are updated accordingly (here, $\beta_{x}$ and $\beta_{y}$ are the participation weights, similar to the $\alpha$ terms of Algorithm \ref{algo_3}). Apart from that, we also use the cycle consistency and identity losses in CLOT-GAN-VC, which are kept similar to FID-RPRGAN-VC \cite{FID-RPR-GAN-VC}.
\par
The proposed collective learning mechanism, integrated with OT loss, is inspired by the core principles of multiplayer games \cite{Multi-player-game} in multi-agent systems \cite{Multi-agent}. In this work, the OT loss (or associated cost matrix) acts as a payoff matrix, where each agent (i.e., discriminator) collaborates based on their participation weights, reflecting their contributions to the loss computation. This results in a collective learning process across the system. The discriminator OT loss implicitly assists the generator by offering highly significant feedback on its generated mel-spectrogram, utilizing the multi-discriminator learning approach. This enables the generator to efficiently learn from its errors and produce mel-spectrograms that are more likely to match the target domain.
\section{Experimental Design} \label{exp-setup}
\subsection{Dataset Description and Training Details}
In this work, the performance evaluation of each model is conducted on VCC 2018 \cite{vcc-2018}, VCTK \cite{CSTR-VCTK}, and CMU Arctic \cite{CMU-Arctic} speech dataset. The CMU-Arctic dataset is also considered in non-parallel settings \cite{cmunp} considering disjoint utterances. The speakers considered for the VCC 2018 dataset are VCC2-TM1/SM3/TF1/SF3, for the VCTK dataset P-229F2/304M2/306F1/334M1, and for the CMU Arctic dataset cmu-us-bld/-rms/-clb/-slt-Arctic. The training, validation and evaluation (test) sets for each dataset comprised of $81$, $35$, $25$ samples, respectively. The objective of using comparatively fewer samples is to evaluate the model's adaptability for low-resource datasets. For training the CLOT-GAN-VC model, the standard \textit{adam} optimizer \cite{adam} is used with a learning rate $1\times 10^{-4}$. The CLOT-GAN-VC model is trained for $1000$ epochs with mini-batch size $1$ and uses input mel-spectrograms of size $2\times80\times64$. Speech reconstruction is performed using a pre-trained MelGAN vocoder \cite{melgan}. 
\par
The proposed model is implemented on a Dell Precision 7820 workstation featuring an Intel Xeon Gold 5215 2.5GHz processor and Nvidia 16GB Quadro RTX5000 graphics. All experiments in this study were carried out using PyTorch 1.1.2 and NumPy 1.19.5.
\section{Results and Discussion}
\label{results}  
\subsection{Objective Evaluation}
The average MCD and MSD values of the proposed CLOT-GAN-VC and the considered GAN-based VC model-generated speech samples are provided in Table \ref{VCC2018-CMU-DATASET}, for each of the three datasets{\footnote{The code implementation and the generated speech samples are available here \url{https://shorturl.at/7X00d}, \url{https://shorturl.at/2BMAj}, respectively}. 
\begin{table}[t]
\centering
\caption{MCD, MSD values for VCC 2018, CMU-Arctic and CSTR-VCTK datasets} 
\label{VCC2018-CMU-DATASET}
\resizebox{0.94\linewidth}{!}{
\begin{tabular}{lccccc}
\hline
\multicolumn{1}{c}{\textbf{Dataset}} & \multicolumn{1}{c}{\textbf{Models}} & \multicolumn{1}{c}{\textbf{M-M}} & \multicolumn{1}{c}{\textbf{F-F}} & \multicolumn{1}{c}{\textbf{M-F}} & \multicolumn{1}{c}{\textbf{F-M}} \\ 
\hline
\multicolumn{6}{c}{\textbf{MCD ($\downarrow$)}} \\ 
\hline
\multicolumn{1}{c}{\multirow{3}{*}{VCC 2018}} & \multicolumn{1}{c}{\textbf{CLOT-GAN-VC}} & 
\multicolumn{1}{c}{\textbf{6.67}} & \multicolumn{1}{c}{\textbf{6.54}} & \multicolumn{1}{c}{\textbf{7.56}} & \textbf{6.69} \\ 
\multicolumn{1}{c}{} & \multicolumn{1}{c}{MaskCycleGAN-VC} & \multicolumn{1}{c}{6.77} & \multicolumn{1}{c}{7.37} & \multicolumn{1}{c}{7.64} & 6.73 \\ 
\multicolumn{1}{c}{} & \multicolumn{1}{c}{MelGAN-VC} & \multicolumn{1}{c}{8.23} & \multicolumn{1}{c}{8.54} & \multicolumn{1}{c}{8.37} & 8.61 \\ \hline
\multicolumn{1}{c}{\multirow{3}{*}{CMU-Arctic}} & \multicolumn{1}{c}{\textbf{CLOT-GAN-VC}} & \multicolumn{1}{c}{\textbf{7.12}} & \multicolumn{1}{c}{8.66} & \multicolumn{1}{c}{\textbf{8.11}} & \textbf{7.91} \\ 
\multicolumn{1}{c}{} & \multicolumn{1}{c}{MaskCycleGAN-VC} & \multicolumn{1}{c}{\textbf{7.12}} & \multicolumn{1}{c}{\textbf{7.81}} & \multicolumn{1}{c}{8.20} & 8.07 \\ 
\multicolumn{1}{c}{} & \multicolumn{1}{c}{MelGAN-VC} & \multicolumn{1}{c}{8.53} & \multicolumn{1}{c}{8.49} & \multicolumn{1}{c}{8.46} & 8.61 \\ \hline
\multicolumn{1}{c}{\multirow{3}{*}{CSTR-VCTK}} & \multicolumn{1}{c}{\textbf{CLOT-GAN-VC}} & \multicolumn{1}{c}{\textbf{4.60}} & \multicolumn{1}{c}{\textbf{7.73}} & \multicolumn{1}{c}{\textbf{4.97}} & {5.86} \\ 
\multicolumn{1}{c}{} & \multicolumn{1}{c}{MaskCycleGAN-VC} & \multicolumn{1}{c}{4.85} & \multicolumn{1}{c}{7.93} & \multicolumn{1}{c}{5.28} & 5.92 \\ 
\multicolumn{1}{c}{} & \multicolumn{1}{c}{MelGAN-VC} & \multicolumn{1}{c}{4.62} & \multicolumn{1}{c}{9.42} & \multicolumn{1}{c}{6.73} & \textbf{5.26} \\ \hline
\multicolumn{6}{c}{\textbf{MSD ($\downarrow$)}} \\ \hline
\multicolumn{1}{c}{\multirow{3}{*}{VCC 2018}} & \multicolumn{1}{c}{\textbf{CLOT-GAN-VC}} & \multicolumn{1}{c}{\textbf{1.21}} & \multicolumn{1}{c}{\textbf{1.15}} & \multicolumn{1}{c}{\textbf{1.42}} & \textbf{1.21} \\ 
\multicolumn{1}{c}{} & \multicolumn{1}{c}{MaskCycleGAN-VC} & \multicolumn{1}{c}{\textbf{1.17}} & \multicolumn{1}{c}{1.18} & \multicolumn{1}{c}{1.50} & 1.24 \\
\multicolumn{1}{c}{} & \multicolumn{1}{c}{MelGAN-VC} & \multicolumn{1}{c}{1.53} & \multicolumn{1}{c}{1.55} & \multicolumn{1}{c}{1.49} & 1.57 \\ \hline
\multicolumn{1}{c}{\multirow{3}{*}{CMU-Arctic}} & \multicolumn{1}{c}{\textbf{CLOT-GAN-VC}} & \multicolumn{1}{c}{\textbf{1.15}} & \multicolumn{1}{c}{\textbf{1.16}} & \multicolumn{1}{c}{1.41} & \textbf{1.23} \\ 
\multicolumn{1}{c}{} & \multicolumn{1}{c}{MaskCycleGAN-VC} & \multicolumn{1}{c}{1.18} & \multicolumn{1}{c}{1.19} & \multicolumn{1}{c}{\textbf{1.32}} & 1.29 \\ 
\multicolumn{1}{c}{} & \multicolumn{1}{c}{MelGAN-VC} & \multicolumn{1}{c}{1.57} & \multicolumn{1}{c}{1.50} & \multicolumn{1}{c}{1.55} & 1.55 \\ \hline
\multicolumn{1}{c}{\multirow{3}{*}{CSTR-VCTK}} & \multicolumn{1}{c}{\textbf{CLOT-GAN-VC}} & \multicolumn{1}{c}{1.40} & \multicolumn{1}{c}{1.63} & \multicolumn{1}{c}{\textbf{1.37}} & \textbf{1.60} \\ 
\multicolumn{1}{c}{} & \multicolumn{1}{c}{MaskCycleGAN-VC} & \multicolumn{1}{c}{\textbf{1.38}} & \multicolumn{1}{c}{\textbf{1.62}} & \multicolumn{1}{c}{1.48} & 1.68 \\ 
\multicolumn{1}{c}{} & \multicolumn{1}{c}{MelGAN-VC} & \multicolumn{1}{c}{1.66} & \multicolumn{1}{c}{1.63} & \multicolumn{1}{c}{1.88} & 1.72 \\ \hline
\end{tabular}}
\end{table}
\begin{figure*}[t]
    \centering
    \includegraphics[height=10.5cm, width=17.5cm]{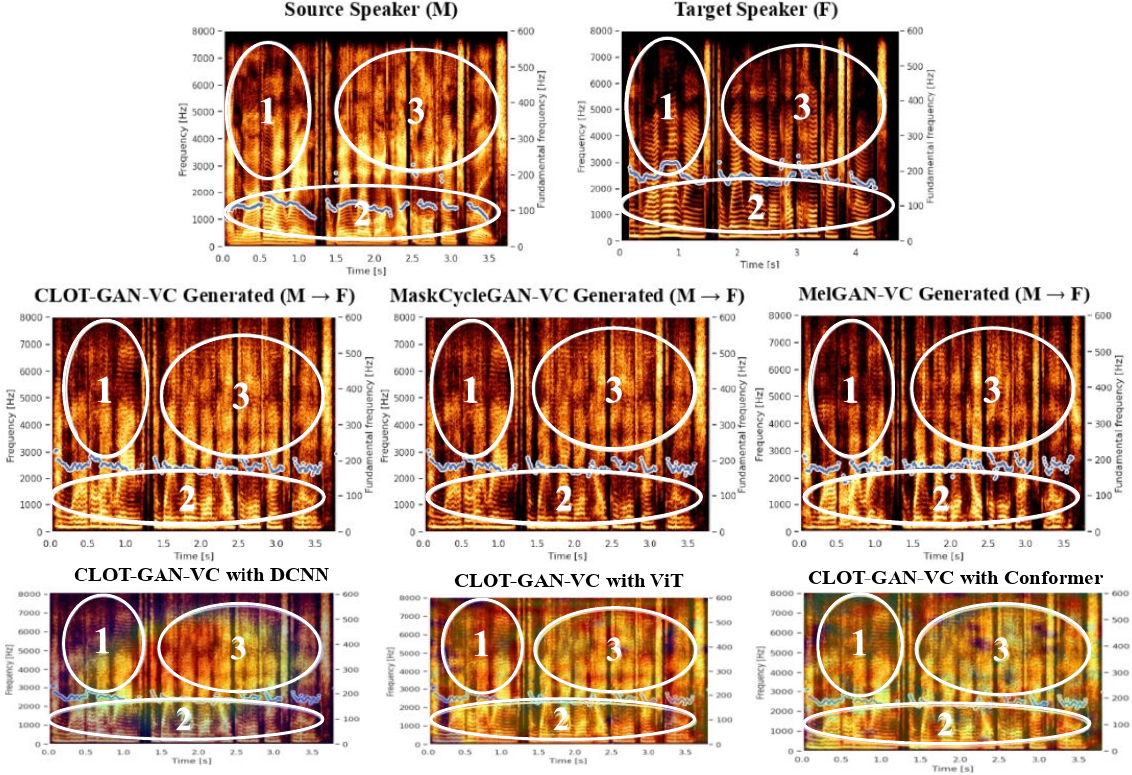}
    \caption{Mel-spectrograms (M to F VC), generated using CLOT-GAN-VC, MaskCycleGAN-VC, and MelGAN-VC, along with Grad-CAM-based visualization for each discriminator setting}
    \label{fig:Mel-spec}
\end{figure*}
The samples for MaskCycleGAN-VC and MelGAN-VC are re-generated for comparison with CLOT-GAN-VC. As indicated in Table \ref{VCC2018-CMU-DATASET}, for both intra and inter-gender VC, the MCD values achieved by CLOT-GAN-VC are lower than those of MaskCycleGAN-VC and MelGAN-VC, except for F-F in CMU-Arctic and F-M in CSTR-VCTK. The difference is more noticeable in the case of any source (M/F) targeting female (F) VC. Considering the average MCD values, the performance improvement is $1.57\%$ and $12.54\%$ in terms of intra-gender VC, and $1.93\%$ and $9.82\%$ in terms of inter-gender VC, w.r.t MaskCycleGAN-VC and MelGAN-VC, respectively. Likewise, based on Table \ref{VCC2018-CMU-DATASET}, it is also evident that the proposed model outperformed the considered SOTA VC models in terms of MSD. However, in the case of intra-gender VC for the CSTR-VCTK dataset, it is evident that MaskCycleGAN-VC exhibited better performance. The average percentage of improvement for intra-gender VC w.r.t MaskCycleGAN-VC and MelGAN-VC is $0.35\%$ and $18.63\%$. Meanwhile, CLOT-GAN-VC also performed significantly better for inter-gender VC than the SOTA VC models considering all the datasets (except CMU-Arctic M-F VC). Apart from the objective evaluation metrics, Fig. \ref{fig:Mel-spec} visually compares the generated speech samples corresponding to each of the models. As depicted in Fig. \ref{fig:Mel-spec}, each mel-spectrogram is divided into three regions (marked as 1,2,3), distinguished by frequency components. Furthermore, as shown in Fig. \ref{fig:Mel-spec}, the energy in both the higher and lower frequency regions of the source mel-spectrogram is noticeably higher than in the target mel-spectrogram. This energy, along with the pitch contour, collectively represents the speech dynamics, while the lower formant patterns reflect the content-related information. Hence, the primary objective of this work is to preserve the utterance-level information while modifying the identity-related aspects. It is evident from Fig. \ref{fig:Mel-spec} that CLOT-GAN-VC efficiently captures the pitch contour of the target speaker, along with the formant patterns that represent the speech content of the input speech (source speaker). Fig. \ref{fig:Mel-spec} also depicts the Grad-CAM \cite{GRAD-CAM} visualizations of the mel-spectrograms for each of the considered discriminators, highlighting the regions that become more prominent for each setting. 
\par
We also report the results of an ablation study in Table \ref{ABLTATION-DATASET} considering the VCC 2018 dataset. 
\begin{table}[t]
\centering
\caption{MCD and MSD values for the ablation study } 
\label{ABLTATION-DATASET}
\resizebox{0.8\linewidth}{!}{
\begin{tabular}{ccccc}
\hline
\textbf{Models} & \textbf{M-M} & \textbf{F-F} & \textbf{M-F} & \textbf{F-M} \\ \hline
\multicolumn{5}{c}{\textbf{MCD ($\downarrow$)}} \\ \hline
CLOT-GAN-VC & \textbf{6.67} & \textbf{6.54} & \textbf{7.56} & \textbf{6.69} \\ 
CLOT-GAN-VC(1) & 6.97 & 7.42 & 7.78 & 6.93 \\ 
CLOT-GAN-VC(2) & 7.07 & 6.75 & 7.64 & 7.22 \\ 
CLOT-GAN-VC(3) & 7.89 & 6.88 & 7.78 & 7.18 \\ \hline
\multicolumn{5}{c}{\textbf{MSD ($\downarrow$)}} \\ \hline
CLOT-GAN-VC & \textbf{1.21} & \textbf{1.15} & \textbf{1.42} & \textbf{1.21} \\ 
CLOT-GAN-VC(1) & 1.21 & 1.18 & 1.50 & 1.44 \\ 
CLOT-GAN-VC(2) & 1.62 & 1.44 & 1.45 & 1.44 \\ 
CLOT-GAN-VC(3) & 1.61 & 1.41 & 1.44 & 1.43 \\ \hline
\end{tabular}}
\end{table}
In Table \ref{ABLTATION-DATASET}, CLOT-GAN-VC(1) indicates the CLOT-GAN-VC model without multiple discriminators (replaced by a single patchGAN discriminator), CLOT-GAN-VC(2) indicates the proposed model without the weighted average operation (replaced by simple average to obtain participation weights), and CLOT-GAN-VC(3) denotes the proposed model without OT loss (replaced by $L_{2}$ loss). 
\par
In Table \ref{ABLTATION-DATASET}, the MCD value of the CLOT-GAN-VC(1) is comparatively lower than the CLOT-GAN-VC(2) and CLOT-GAN-VC(3) models for M-M VC. Based on the MCD values obtained for M-M VC, the models CLOT-GAN-VC(1), CLOT-GAN-VC(2) and CLOT-GAN-VC(3) can be ranked as rank-1, rank-2 and rank-3 respectively. This implies that removing OT loss significantly degraded the results in terms of MCD value. This indicates the effectiveness of  OT loss in M-M VC. Whereas in terms of MSD value, though CLOT-GAN-VC(1) obtained rank-1, the ranks of CLOT-GAN-VC(2) and CLOT-GAN-VC(3) swapped from their respective MCD ranks (for both M-M and F-F VC). However, the difference between MSD values of CLOT-GAN-VC(2) and CLOT-GAN-VC(3) for M-M and F-F VC is very small (0.01 and 0.03, respectively). From this, it is evident that in the case of intra-gender VC, both the OT loss and the weighted average mechanism significantly impact the overall performance of the CLOT-GAN-VC model. In contrast, it is also noticeable that the ranks of the respective CLOT-GAN-VC settings based on MCD values obtained for F-F VC differ from their M-M counterpart. From the outcome of the ablation study conducted for F-F VC (in terms of MCD), it is observable that removing the multi-discriminator from CLOT-GAN-VC highly impacted the results compared to removing other components from the original proposed model.
\par
On the other hand, for inter-gender VC, the ranks of CLOT-GAN-VC(1), CLOT-GAN-VC(2), and CLOT-GAN-VC(3) are different for both M-F and F-M VC (in terms of MCD). For M-F VC the CLOT-GAN-VC(1)/CLOT-GAN-VC(3) obtained same rank (rank-2), whereas for F-M VC CLOT-GAN-VC(2) and CLOT-GAN-VC(3) obtained rank-3 and rank-2 respectively. This suggests that removing OT loss significantly impacts the overall model's performance in both cases (considering MCD). However, in terms of MSD, the ranks of all the CLOT-GAN-VC settings are in the same order (i.e., CLOT-GAN-VC(1)/-VC(2)/-VC(3) ranked as rank-3/-2/-1 respectively). This reveals that each component exerts a comparable influence in the context of inter-gender VC (taking MSD into account), with the multi-discriminator being the most substantial.

\subsection{Subjective Evaluation}
In this work, we collected subjective evaluations of the generated speech samples to assess the naturalness based on MOS \cite{mos}. A total of 17 volunteers participated in the evaluation process. To obtain the MOS values, generated speech samples were collected from each dataset. The volunteers rated randomly chosen speech samples for each VC category (intra-/inter-gender) using a scale ranging from 1 to 5 (based on naturalness){\footnote{The MOS collection videos, along with all other relevant details are available here \url{https://tinyurl.com/959y6h4c}}}, following the standard MOS calculation method \cite{mos}. We provided the MOS values with $95\%$ confidence intervals in Table \ref{MOS} for VCC 2018, CMU-Arctic, and CSTR-VCTK datasets. 
\begin{table}[t]
\centering
\small
\caption{MOS $(\uparrow)$ with $95\%$ confidence intervals} 
\label{MOS}
\resizebox{0.99\linewidth}{!}{\begin{tabular}{cccccc}
\hline
\textbf{Dataset} & \textbf{Models} & \textbf{M-M} & \textbf{F-F} & \textbf{M-F} & \textbf{F-M} \\ 
\hline
\multirow{3}{*}{VCC 2018} & \textbf{CLOT-GAN-VC} & \textbf{3.57$\pm$0.53} & \textbf{3.73$\pm$0.57} & \textbf{2.73$\pm$0.21} & \textbf{2.84$\pm$0.38} \\  
 & MaskCycleGAN-VC & 3.47$\pm$0.50 & 3.26$\pm$0.44 & 2.53$\pm$0.19 & 2.68$\pm$0.33 \\
 & MelGAN-VC & 2.10$\pm$0.93 & 2.10$\pm$0.99 & 2.21$\pm$0.13 & 2.59$\pm$0.04 \\\hline
\multirow{3}{*}{CMU-Arctic} & \textbf{CLOT-GAN-VC} & \textbf{3.22$\pm$.09} & \textbf{4.11$\pm$0.45} & \textbf{3.05$\pm$0.05} & \textbf{2.88$\pm$0.97} \\
 & MaskCycleGAN-VC & 3.16$\pm$0.03 & 3.50$\pm$0.24 & 2.72$\pm$0.01 & 2.61$\pm$0.18 \\ 
 & MelGAN-VC & 2.72$\pm$0.82 & 2.88$\pm$0.83 & 2.38$\pm$0.19 & 2.11$\pm$0.07 \\\hline
 \multirow{3}{*}{CSTR-VCTK} & \textbf{CLOT-GAN-VC} & \textbf{3.63$\pm$0.57} & \textbf{3.21$\pm$0.43} & \textbf{3.36$\pm$0.38} & \textbf{3.10$\pm$0.22} \\
 & MaskCycleGAN-VC & 3.52$\pm$0.54 &  2.94$\pm$0.26 & 3.31$\pm$0.45 & 2.94$\pm$0.44 \\ 
 & MelGAN-VC & 2.84$\pm$0.57 &  2.63$\pm$0.83 &  2.05$\pm$0.91 & 2.00$\pm$0.88 \\\hline
 \vspace{-0.67cm}
\end{tabular}}
\end{table}
It is observed from Table \ref{MOS} that the MOS values of the proposed model are significantly higher than the considered models. Hence, the subjective evaluation results indicate that the proposed model outperforms the other considered VC models.

\section{Conclusion} \label{conclusion}
This study presents a novel approach that leverages a collaborative learning mechanism within a GAN model for VC. The GAN utilizes multiple discriminators with diverse deep learning architectures and integrates an OT loss function based on the Sinkhorn algorithm. The multi-discriminator framework is designed to collaboratively capture the formant distribution in mel-spectrograms by effectively identifying differences between source and target feature distributions through OT loss. Experimental results show that the converted speech samples from our proposed CLOT-GAN-VC model outperform others. Additionally, investigating the model's effectiveness for voice conversion in low-resource languages could serve as a promising direction for future research.
\bibliographystyle{IEEEtran}
\bibliography{reference.bib}
\end{document}